\documentclass[reprint,amsmath,amssymb,aps,nofootinbib]{revtex4-2}
\usepackage{graphicx}
\usepackage{dcolumn}
\usepackage{bm}
\usepackage{braket}
\usepackage{comment}  
\usepackage{soul}  
\usepackage{color}
\usepackage{afterpage}
\usepackage[colorlinks=true]{hyperref}

\graphicspath{{figures/}}

\begin{document}

\author{Rapha\"el-David Lasseri} 
\affiliation{Centre Borelli, ENS Paris-Saclay, 91190 Gif-sur-Yvette, France}
\affiliation{Magic LEMP, 94110 Arcueil, France}

\author{David Regnier} 
\email{david.regnier@cea.fr}
\affiliation{CEA, DAM, DIF, 91297 Arpajon, France}
\affiliation{Université Paris-Saclay, CEA, Laboratoire Matière en Conditions Extrêmes, 91680 Bruyères-le-Châtel, France}

\author{Mikaël Frosini}
\affiliation{CEA, DEN, IRESNE, DER, SPRC, 13108 Saint-Paul-lès-Durance, France}

\author{Marc Verriere}
\affiliation{Nuclear and Data Theory Group, Nuclear and Chemical Science Division, Lawrence Livermore National Laboratory, Livermore, CA, United States}

\author{Nicolas Schunck}
\affiliation{Nuclear and Data Theory Group, Nuclear and Chemical Science Division, Lawrence Livermore National Laboratory, Livermore, CA, United States}


\title{Generative deep-learning reveals collective variables of Fermionic systems}

\maketitle

%
{\fontsize{13pt}{0}\selectfont
Complex processes ranging from protein folding to nuclear fission often follow a low-dimension reaction path parameterized in terms of a few collective variables.
In nuclear theory, variables related to the shape of the nuclear density in a mean-field picture
are key to describing the large amplitude collective motion of the neutrons and protons. 
Exploring the adiabatic energy landscape spanned by these degrees of freedom reveals the possible reaction channels while simulating the dynamics in 
this reduced space yields their respective probabilities.
Unfortunately, this theoretical framework breaks down whenever the systems 
encounters a quantum phase transition with respect to the collective variables.
Here we propose a generative-deep-learning algorithm capable of building new collective variables highly representative of a nuclear process while ensuring a differentiable mapping to its Fermionic wave function.
Within this collective space, the nucleus can evolve continuously from one of its adiabatic quantum phase to the other at the price of crossing a potential energy barrier.
This approach applies to any Fermionic system described by a single Slater determinant, which encompasses electronic systems described within the density functional theory.
}

Describing the dynamics of Fermionic systems is key to understanding how molecules react to a laser excitation~\cite{gonzalez2020quantum}, study quantum phase transitions in ultracold gases~\cite{bloch2008manybody} or simulate low energy heavy-ion collisions~\cite{simenel2018heavyion}.
For most mesoscopic systems, the exponential growth of the Hilbert space with the number of particles involved makes an exact resolution of this quantum many-body problem unreachable.
Yet, various complex processes such as large amplitude vibrations in atomic clusters or atomic nuclei
emerge as collective behaviors of their constituents~\cite{dinh2017similarities}. 
In such cases, the system's wave function mostly remains within
a low-dimensional sub-manifold of the complete phase-space.
A few collective variables are sufficient 
to parameterize its dynamics.
In this situation, a widespread theoretical approach consists in computing a series of static low-energy Fermionic wave functions
along the collective path~\cite{manzhos2021neural,schlegel2003exploring,schunck2022theory}.
This yields the so-called adiabatic energy landscape that reveals favored configurations or reaction channels.
Going one step further, it is possible to recover a classical or quantum mechanical equation of motion in terms of the collective variables~\cite{hollingsworth2018molecular,verriere2020timedependent}.
Integrating this equation in time gives access to the timescales of the phenomenon along with the probability of populating different final states.
The success of this theoretical framework requires that (i) the collective subspace is representative of the phenomenon under study (ii) the mapping from the collective variable to the quantum state of the system is differentiable. 
However, computing a landscape of adiabatic quantum states that minimizes the energy at a given collective coordinate gives no guarantee that the latter condition is fulfilled.
%

Chemistry and biochemistry possess a large corpus of methods to build collective variables describing the positions of the nuclei along a reaction path~\cite{rogal2021reaction,maeda2015intrinsic,stewart1987semiempirical,denzel2019gaussian,tiwary2013metadynamics}.
A method such as meta-dynamics built on top of \textit{ab-initio} electronic states ensures the continuity of the positions of the nuclei as a function of the collective variable~\cite{bussi2020using,schlegel2003exploring}.
Yet brutal changes in the electronic configurations may still appear and manifest, for instance, as energy cusps at conical intersections~\cite{evenhuis2011scheme}.
A similar situation holds in nuclear physics, where collective variables related to the intrinsic shape of the nuclear density are commonly
used to study low-energy structure, giant resonances, fusion, and fission reactions~\cite{bender2008going}.
Within such collective space, the clusterization of some nucleons~\cite{girod2013alphaparticle,ebran2020ensurematha} 
or the discontinuous changes in the shape of a fissioning nucleus appear as first-order quantum phase transitions (QPT)~\cite{dubray2012numerical,regnier2019asymmetric}.
In several studies, the adiabatic energy landscape was pragmatically smoothed out
in order to be able to simulate the dynamics nonetheless~\cite{ho1996general,hurd2010trajectory,evenhuis2011scheme,galvao2016modeling}.
Another approach consists in going beyond the adiabatic approximation at the price of greater complexity and numerical cost~\cite{curchod2018initio,varandas2009simple,galvao2016modeling,bernard2011microscopic,dietrich2010microscopic,zhao2019microscopic}.
Recently, Lau \textit{et al.} proposed a novel method that avoids these QPT by locally modifying the collective path~\cite{lau2022smoothing}.
Unfortunately, this approach requires additional mean-field calculations, and its success is not guaranteed for all QPT.

In parallel with these efforts, the fast development of dimension reduction algorithms in the field of artificial intelligence opens new alternatives to the design of collective variables.
In the context of molecular systems, small dimensional representations of the positions
of nuclei were obtained with the help of genetic algorithms~\cite{ma2005automatic},
principal component analysis, Isomaps, Sketch-maps or diffusion maps~\cite{rogal2021reaction}
but also autoencoders and variational autoencoders~\cite{baima2022capabilities,belkacemi2022chasing}.
An additional difficulty arises when the collective variables directly map Fermionic wave functions that belong to a non-Euclidean space and possess peculiar symmetries~\cite{verriere2022building}.
Yet both fields of neural networks representation of Fermionic states~\cite{carleo2018constructing,carleo2019machine,schutt2019unifying,companysfranzke2022excited,cassella2023discovering} and neural networks
for Riemannian geometries~\cite{asif2021graph,zhang2018grassmannian,chakraborty2022manifoldnet} are quickly progressing.

Based on these achievements, we show that it is
possible to build differentiable collective variables directly mapping Slater determinants and demonstrate its ability to better describe the dynamics close to a QPT in nuclear physics.

\section{Quantum phase transition in the $^{16}$O atomic nucleus}

We consider the case of the atomic nucleus modeled as an ensemble of point-like neutrons and protons.
Among the thousands of nuclear systems of interest, we choose to demonstrate our method on a nucleus of $^{16}$O. 
Due to its closed shell nature, its superfluidity is negligible, which makes it similar to a large number of electronic systems encountered in chemistry and condensed matter.
The Hartree-Fock ground state of $^{16}$O, obtained with a Machleidt \textit{ab-initio} nucleon-nucleon interaction~\cite{machleidt2016chiral}, possesses a spherical one-body density.
As standard in nuclear physics, we study the static response of this nucleus to an external force field constraining its elongation~\cite{bender2003selfconsistent}.
Here, the collective variable is the expectation value of the intrinsic quadrupole moment $Q_{20}=\langle \hat{Q}_{20}\rangle$ of the nucleus.
Solving the constrained Hartree-Fock equation for 200 regularly spaced values of the quadrupole moment yields 
the adiabatic collective path. 

\begin{figure*}[ht]
    \centering
    \includegraphics[width=0.8\textwidth]{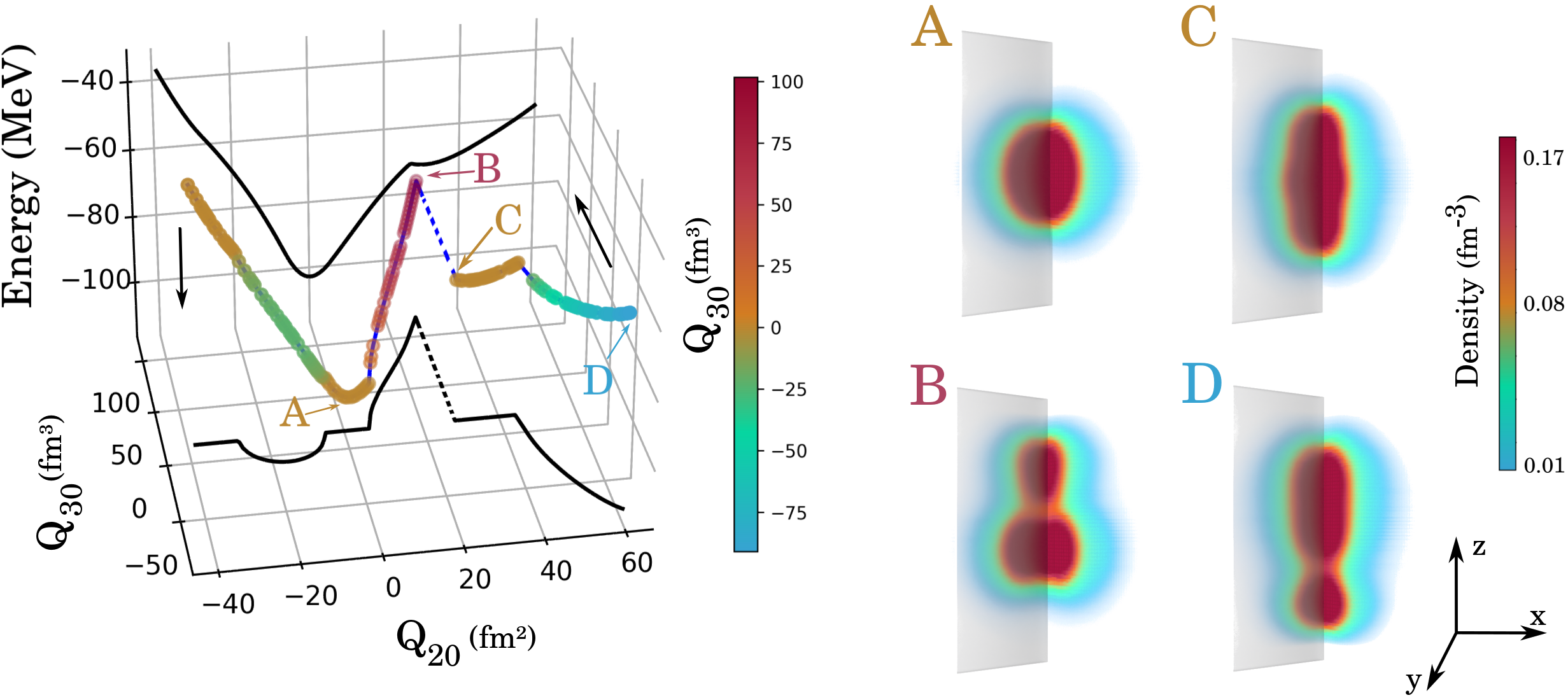}
    \caption{Left Panel: The quadrupole-constrained Hartree-Fock path of $^{16}$O plotted in the ($Q_{20}$, $Q_{30}$, energy) space. The dark continuous lines represent its projections on the ($Q_{20}$, Energy) and ($Q_{20}$,$Q_{30}$) planes. The color of the path corresponds to its $Q_{30}$ values. 
    The ground state of the system is the spherical configuration A.
    The points B and C are separated by a first-order QPT highlighted by dashed lines.
    The point D shows a typical high quadrupole deformation.
    Right Panel: Local nucleonic densities for the indicated points of the quadrupole-constrained Hartree-Fock path.}
    \label{fig:o16_adiabatic_pes}
\end{figure*}

The Figure~\ref{fig:o16_adiabatic_pes} illustrates this collective path in terms of the quadrupole and octupole moments of the 
solutions along with their Hartree-Fock energy.
%
In spite of the continuous behavior exhibited by the energy as a function of the quadrupole moment, a sudden transition occurs in the system, shifting discontinuously from the pear-shaped configuration B to the axially-symmetric configuration C.
This abrupt change in the quantum state of the nucleus is a first-order quantum phase transition. 
The Fubini-Study distance~\cite{carollo2020geometry} between the quantum states B and C is 
two orders of magnitude higher
than typical between other neighboring states.
This phase transition can be understood as having a geometric nature, where the order parameter is represented by the nucleus's octupole moment, denoted as $Q_{30}=\langle \hat{Q}_{30}\rangle$.
In addition to this first-order QPT, there are four other points in the collective space where the derivative of the octupole moment also exhibits discontinuities.
These points correspond to second-order QPTs of the system, where the first derivative of the quantum state with respect to the collective variable $Q_{20}$ is discontinuous.

The presence of these QPTs in the collective space poses a significant challenge when attempting to describe the system's dynamics. The collective inertia, which plays a crucial role in simulating large-amplitude collective dynamics and is a key driver of spontaneous fission half-lives \cite{sadhukhan2020microscopic}, 
becomes ill-defined in the vicinity of the QPTs~\cite{verriere2020timedependent}.
The practical approach to regularize
these inertia commonly employed in state-of-the-art fission dynamics calculations would likely result in an uncontrolled overestimation of the decay rate from the prolate metastable state at $Q_{20}\simeq 30$ fm$^2$ towards the ground state.

\section{Building an alternative collective variable}

In this study, we use a set of adiabatic $Q_{20}$-constrained quantum states to learn a continuous representation of the nuclear collective space.
The data set represented by Fig.~\ref{fig:o16_adiabatic_pes} consists of 200 quantum states for $^{16}$O.
As usual in machine learning, we split the data randomly into a 70\% training set, 20\% validation set and 10\% test set.
The calibration of our dimension reduction algorithm involves the training and validation sets only while its final predictive power is estimated from the test set. As shown later, this sampling has a low impact on the final results.
Each data point stands for a tensor product of one time-reversal-invariant Slater determinant for the 8 neutrons and another one for the 8 protons.
The single-particle orbitals are expanded on a truncated basis of 330 eigenstates of an axially-symmetric harmonic oscillator Hamiltonian. The coefficients of this expansion fully define a nuclear state.
Yet, this mapping is not bijective 
since a unitary transform of these coefficients may parameterize the same Slater determinant.
More specifically, the classes of normalized Slater determinants which are equal up to a phase, often named Slater rays, form a set that is isomorphic to the Grassmannian manifold, a non-Euclidean geometry~\cite{aoto2020calculating}.

In order to leverage standard dimension reduction techniques on such objects, we 
first 
need to 
establish a differential mapping from a vector of real coordinates to the Slater determinants of our data set.
In this paper, we explore two alternatives choices for this mapping.
A first approach parameterizes the class of Slater determinants up to a phase by 
the elements of the upper triangular part of their one-body density matrices expressed in the harmonic oscillator basis.
Within the one-body space considered, this mapping requires 108240 real numbers for the parameterization of one $^{16}$O state.
It is simple to implement but does not
yield a one-to-one mapping, as some values for the density matrix may not
represent any Slater determinant.
A more complex but more robust approach leverages the Thouless theorem to parameterize our finite set of Slater determinants.
In this case, we first compute a pivotal Slater determinant as the Karcher mean of our training dataset~\cite{huper2013averaginga,JMLR:v17:16-177}.
This new state minimizes its average geodesic distance to the training dataset in the Grassmannian manifold.
We then parameterize any Slater determinant relative to this pivotal point $|\phi_0\rangle$ as
\begin{equation}
 | \phi(N,\theta,Z) \rangle = Ne^{i\theta} 
 \text{exp}\left( 
 \sum_{mi}Z_{mi} a^{\dagger}_ma_i
 \right)
 |\phi_0\rangle.
\end{equation}
The positive real number $N$ and the angle $\theta$ parameterize the norm and phase of the state while the real matrix elements $Z_{mi}$ contain the information
on the Slater ray itself. 
In most applications, only the $Z$ coordinates present interest as a phase and norm convention could be chosen arbitrarily.
The sum runs over particle (resp. hole) orbitals of the pivotal Slater determinant for the index $m$ (resp. $i$).
This method ensures a one-to-one differentiable mapping as long as the pivotal point has a non-vanishing overlap with all the Slater determinant 
in the set.
Describing an $^{16}$O elongated state requires two vectors of 1304 real parameters, one associated with the neutron $Z$ matrix and the other one with the protons.
In this study, both the density matrix and Thouless representations lead to similar conclusions.
In the following we present only
results obtained with the Thouless representation.

The next step to building a collective variable consists in finding a low-dimension representation of the training data set.
To do so, we combine a principal component analysis (PCA) with a variational autoencoder (VAE) in a similar manner as in \cite{baima2022capabilities}.
The former performs a nearly lossless compression/decompression from the pairs of the neutrons and protons $Z$ coordinates to an intermediate space $\mathcal{S}_{\text{PCA}}$.
Selecting the first 20 principal components enables a root mean square error lower than $10^{-7}$ for the $Z$ matrices in the training set.
This linear transformation already gives a significant dimension reduction at a relatively low numerical cost.
Finally, we train a variational auto-encoder to further reduce the dimension from the training set in $\mathcal{S}_{\text{PCA}}$ to a final latent space of dimension one.
We specifically choose a VAE as opposed to an autoencoder so to regularize the latent space
and improve the generative quality of the algorithm.
The details of the VAE architecture selected and the process of hyper-parameter tuning are reported in Ref.\ref{sec:vae_training}.
After training, the PCA decomposition followed by the encoder of the VAE maps any Thouless representation (1) into a single latent variable $\lambda$.
In the same way, any $\lambda$ value passed through the VAE decoder, the inverse PCA transformation and the inverse of the Thouless transformation gives a
well defined Slater determinant of $^{16}$O.
Finally, using differentiable activation functions for the variational auto-encoder ensures the differentiable nature of the encoder and decoder.
This makes $\lambda$ a new differentiable collective variable.
Its ability to span at least the same states as the previous $Q_{20}$ representation is directly related to the convergence of our training step.
%
Here, the root mean square reconstruction error on the energy is below 0.3\% of the ground state energy for the training, validation and test sets.
Similarly the root mean square errors on the quadrupole and octupole moments are below 0.01\% of their respective variation range.
This demonstrates the quality of the training.

\section{Deformation properties along the new collective variable}

\begin{figure}[!ht]
\includegraphics[width=0.5\textwidth]{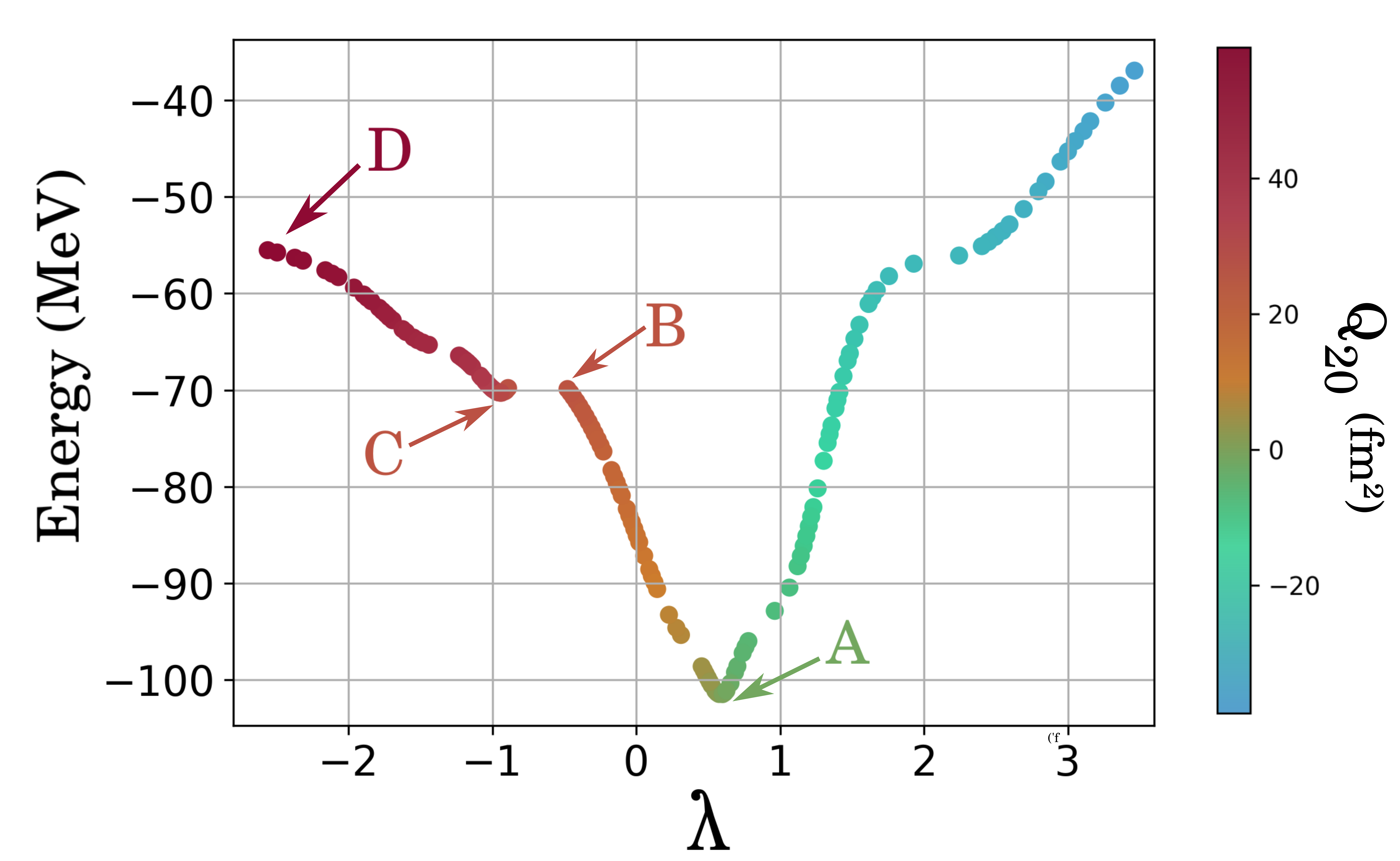}
\caption{Energy landscape in the latent space discovered by the machine learning algorithm.
Points are colored according to the $Q_{20}$ coordinate revealing the monotonous evolution of $Q_{20}$ with $\lambda$. Arrows report the configurations A, B, C, D of Fig.~\ref{fig:o16_adiabatic_pes}.}
\label{fig:latent_space}
\end{figure}

Figure \ref{fig:latent_space} shows the Hartree-Fock energy and the quadrupole moment of $^{16}$O as a function of the $\lambda$ collective variable 
for the quantum states of the training set.
The topology of the energy curve is very close to the $Q_{20}$ representation of Fig.~\ref{fig:o16_adiabatic_pes} up to a change of sign in abscissa.
The $\lambda$ coordinate decreases monotonously with $Q_{20}$ and can be interpreted in a similar way as encoding an information about the elongation of the system.
The $\lambda$ representation clearly separates the peared-shape configuration $B$ from the axially-symmetric one $C$ even though they share the same quadrupole moment.
In addition, small gaps appear at points like $\lambda=-2, Q_{20}=45$ fm$^2$ corresponding to the second order QPTs in Fig.~\ref{fig:o16_adiabatic_pes}.
As a first result, the $\lambda$ collective variable better discriminates the quantum phases of our system.
%

\begin{figure}[!ht]
\includegraphics[width=0.5\textwidth]{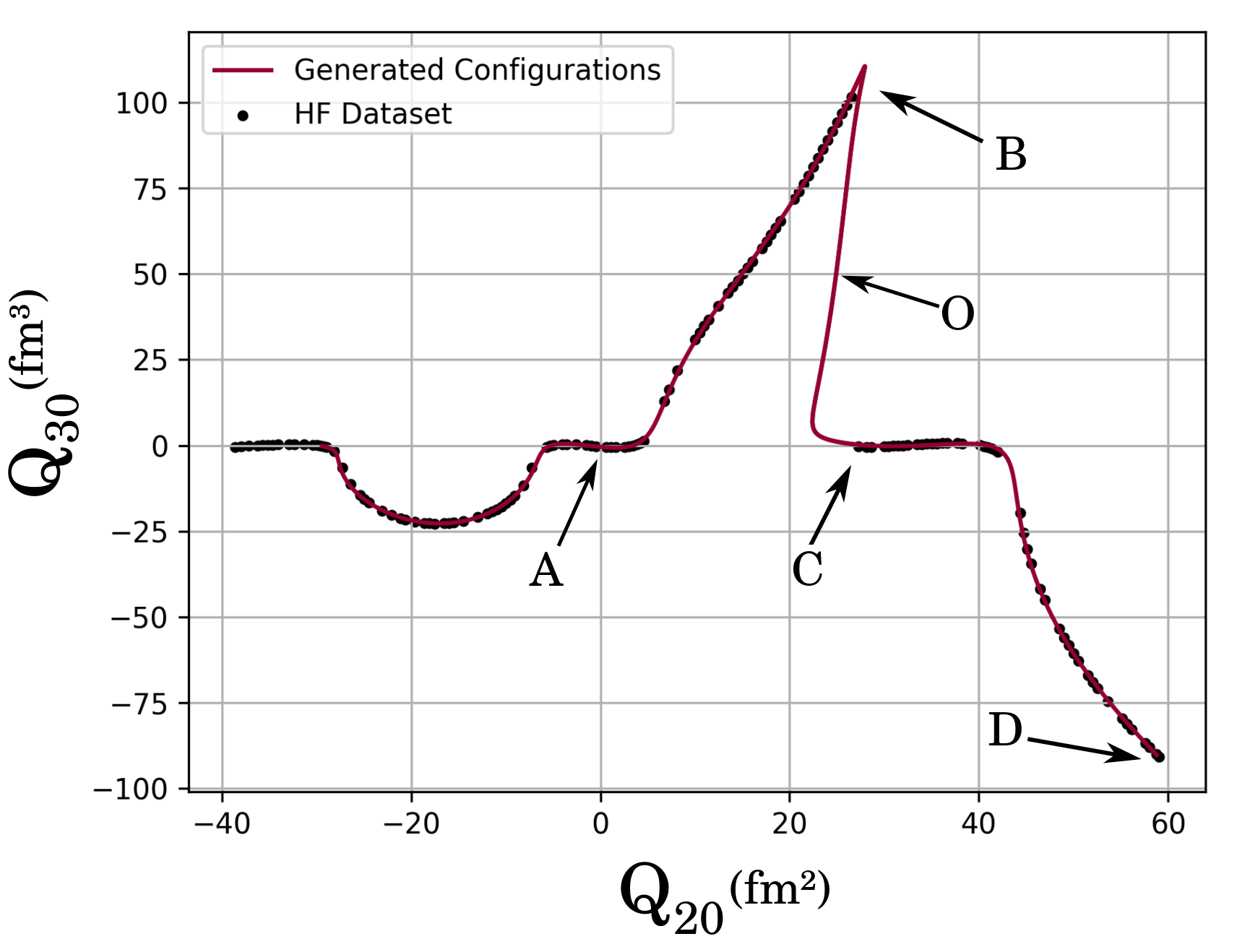}
\includegraphics[width=0.5\textwidth]{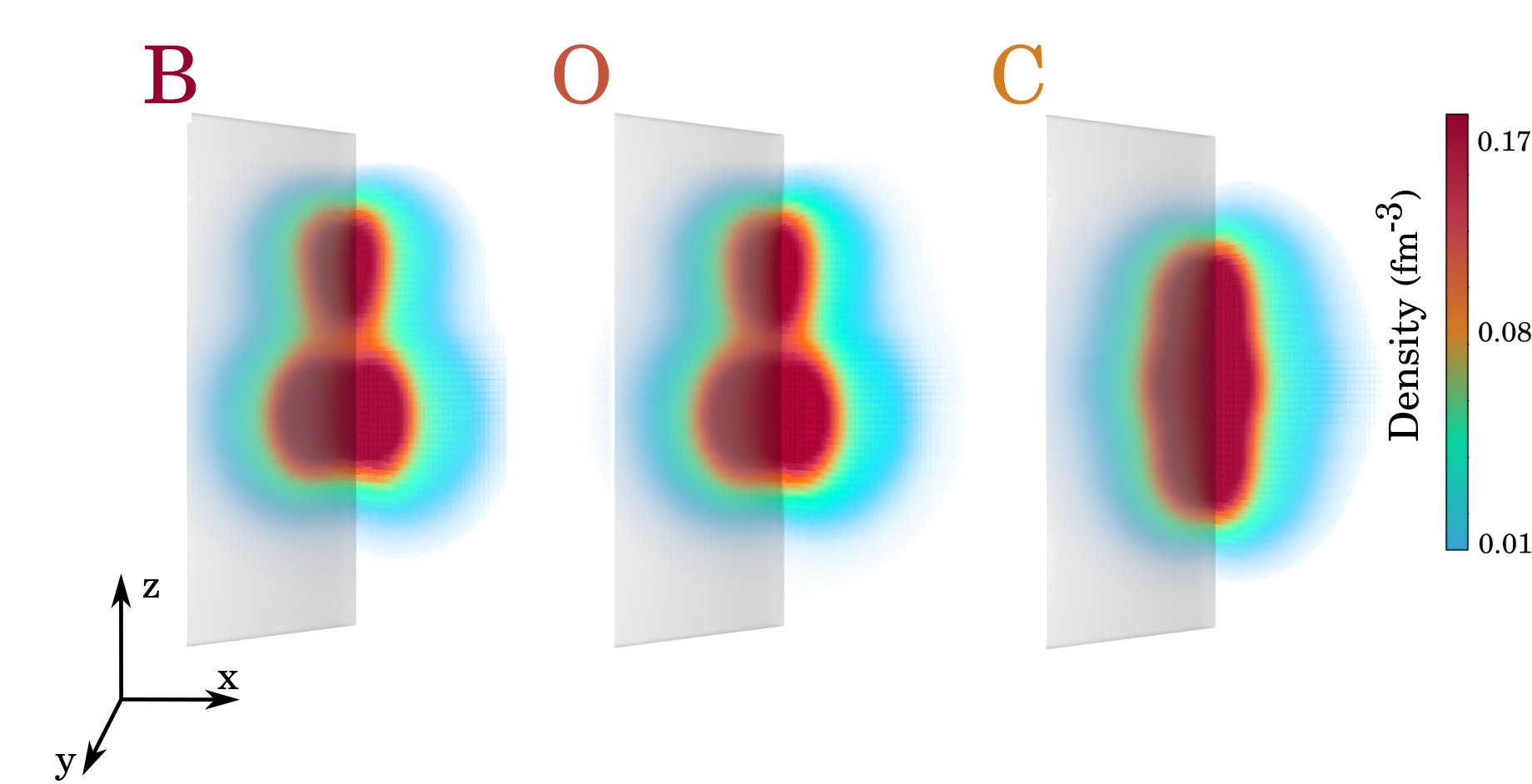}
\caption{Upper panel: Collective path generated by the machine learning algorithm represented in the ($Q_{20},Q_{30}$) space.
Lower panel: Local nucleonic densities for the indicated points. 
The new configuration $O$ corresponds to a $\lambda$ value in between the ones of configurations $B$ and $C$.}
\label{fig:prolongation_densities}
\end{figure}

Going one step further, we can now connect in a differentiable manner the adiabatic phases.
For this, we choose a fine enough mesh in the latent space spanned by the variable $\lambda$ 
and generate for each point of this mesh a new quantum state. 
The numerical cost associated with this step is negligible compared to the one required by a series of new Hartree-Fock calculations.
Figure \ref{fig:prolongation_densities} shows the one-body density and multipole moments along the new collective coordinates.
Their values continuously connect configurations B and C.

\begin{figure}[!ht]
\includegraphics[width=0.5\textwidth]{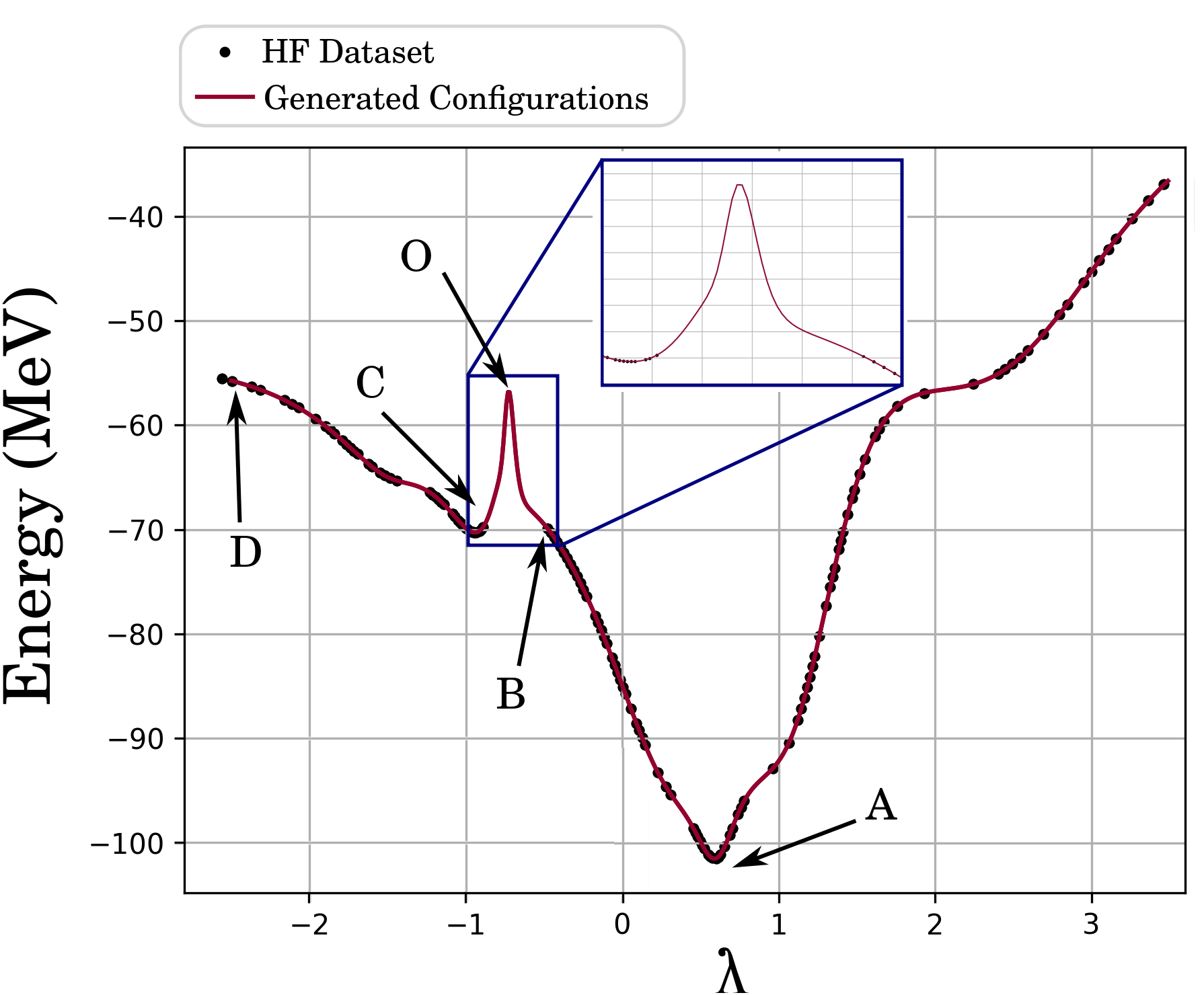}
\caption{Energy landscape in the latent space discovered by the machine learning algorithm.
Points generated from the latent space connect the configurations B and C lying in two different quantum phases
and reveal a new potential barrier.
}
\label{fig:prolongation_energy}
\end{figure}

Finally, figure~\ref{fig:prolongation_energy} shows the Hartree-Fock energy as a function of $\lambda$ 
obtained by (i) choosing a fine $\lambda$-grid (ii) generating quantum states for each $\lambda$ of this grid.
A potential barrier between configurations B and C appears as a striking new feature.
The height of this new barrier is 13.2 MeV which should significantly hinder a dynamical transition
from the prolate configurations to the ground state of the system.
This value depends on the detail of the many-body path from configuration B to C.
Typically, we found that training multiple times the same VAE with different initial weights 
always yields a new potential barrier. Its height varies with a standard deviation of 2.1 MeV (cf. \ref{sec:heights_sensitivity}).
%

\section{Conclusion}

In this study, we present a groundbreaking approach that harnesses the power of 
generative
machine learning for dimensional reduction techniques, enabling the construction of a novel collective variable that accurately represents nuclear processes. 
By ensuring a differentiable parametrization of quantum states, our method generates continuous trajectories between distinct quantum phases, a crucial aspect for estimating the energy required for quantum phase transitions.

Our findings open up new avenues for exploring nuclear dynamics within a comprehensive quantum framework, such as the time-dependent generator coordinate method. 
This advancement holds the potential to predict half-lives of metastable states and the relative probabilities of reaction outcomes.

Moreover, our approach is applicable to a broad range of Fermionic systems represented by Slater determinant states, including the vast domain of electronic systems described through density functional theory.
The adaptability and versatility of neural networks ensure that our method can be extended to accommodate even more complex Fermionic states in the future.

In summary, this study paves the way for a deeper understanding of nuclear processes, electronic systems, and Fermionic states, demonstrating the remarkable potential of integrating machine learning techniques in the realm of quantum physics.


\begin{thebibliography}{54}%
\makeatletter
\providecommand \@ifxundefined [1]{%
 \@ifx{#1\undefined}
}%
\providecommand \@ifnum [1]{%
 \ifnum #1\expandafter \@firstoftwo
 \else \expandafter \@secondoftwo
 \fi
}%
\providecommand \@ifx [1]{%
 \ifx #1\expandafter \@firstoftwo
 \else \expandafter \@secondoftwo
 \fi
}%
\providecommand \natexlab [1]{#1}%
\providecommand \enquote  [1]{``#1''}%
\providecommand \bibnamefont  [1]{#1}%
\providecommand \bibfnamefont [1]{#1}%
\providecommand \citenamefont [1]{#1}%
\providecommand \href@noop [0]{\@secondoftwo}%
\providecommand \href [0]{\begingroup \@sanitize@url \@href}%
\providecommand \@href[1]{\@@startlink{#1}\@@href}%
\providecommand \@@href[1]{\endgroup#1\@@endlink}%
\providecommand \@sanitize@url [0]{\catcode `\\12\catcode `\$12\catcode
  `\&12\catcode `\#12\catcode `\^12\catcode `\_12\catcode `\%12\relax}%
\providecommand \@@startlink[1]{}%
\providecommand \@@endlink[0]{}%
\providecommand \url  [0]{\begingroup\@sanitize@url \@url }%
\providecommand \@url [1]{\endgroup\@href {#1}{\urlprefix }}%
\providecommand \urlprefix  [0]{URL }%
\providecommand \Eprint [0]{\href }%
\providecommand \doibase [0]{http://dx.doi.org/}%
\providecommand \selectlanguage [0]{\@gobble}%
\providecommand \bibinfo  [0]{\@secondoftwo}%
\providecommand \bibfield  [0]{\@secondoftwo}%
\providecommand \translation [1]{[#1]}%
\providecommand \BibitemOpen [0]{}%
\providecommand \bibitemStop [0]{}%
\providecommand \bibitemNoStop [0]{.\EOS\space}%
\providecommand \EOS [0]{\spacefactor3000\relax}%
\providecommand \BibitemShut  [1]{\csname bibitem#1\endcsname}%
\let\auto@bib@innerbib\@empty
\bibitem [{\citenamefont {Gonz{\'a}lez}(2020)}]{gonzalez2020quantum}%
  \BibitemOpen
  \bibfield  {author} {\bibinfo {author} {\bibfnamefont {L.}~\bibnamefont
  {Gonz{\'a}lez}},\ }\href@noop {} {\emph {\bibinfo {title} {{Quantum Chemistry
  and Dynamics of Excited States: Methods and Applications}}}}\ (\bibinfo
  {publisher} {{Wiley-Blackwell}},\ \bibinfo {address} {{Hoboken, NJ}},\
  \bibinfo {year} {2020})\BibitemShut {NoStop}%
\bibitem [{\citenamefont {Bloch}\ \emph {et~al.}(2008)\citenamefont {Bloch},
  \citenamefont {Dalibard},\ and\ \citenamefont {Zwerger}}]{bloch2008manybody}%
  \BibitemOpen
  \bibfield  {author} {\bibinfo {author} {\bibfnamefont {I.}~\bibnamefont
  {Bloch}}, \bibinfo {author} {\bibfnamefont {J.}~\bibnamefont {Dalibard}}, \
  and\ \bibinfo {author} {\bibfnamefont {W.}~\bibnamefont {Zwerger}},\ }\href
  {\doibase 10.1103/RevModPhys.80.885} {\bibfield  {journal} {\bibinfo
  {journal} {Rev. Mod. Phys.}\ }\textbf {\bibinfo {volume} {80}},\ \bibinfo
  {pages} {885} (\bibinfo {year} {2008})}\BibitemShut {NoStop}%
\bibitem [{\citenamefont {Simenel}\ and\ \citenamefont
  {Umar}(2018)}]{simenel2018heavyion}%
  \BibitemOpen
  \bibfield  {author} {\bibinfo {author} {\bibfnamefont {C.}~\bibnamefont
  {Simenel}}\ and\ \bibinfo {author} {\bibfnamefont {A.~S.}\ \bibnamefont
  {Umar}},\ }\href {\doibase 10.1016/j.ppnp.2018.07.002} {\bibfield  {journal}
  {\bibinfo  {journal} {Progress in Particle and Nuclear Physics}\ }\textbf
  {\bibinfo {volume} {103}},\ \bibinfo {pages} {19} (\bibinfo {year}
  {2018})}\BibitemShut {NoStop}%
\bibitem [{\citenamefont {Dinh}\ \emph {et~al.}(2017)\citenamefont {Dinh},
  \citenamefont {Reinhard},\ and\ \citenamefont
  {Suraud}}]{dinh2017similarities}%
  \BibitemOpen
  \bibfield  {author} {\bibinfo {author} {\bibfnamefont {P.~M.}\ \bibnamefont
  {Dinh}}, \bibinfo {author} {\bibfnamefont {P.-G.}\ \bibnamefont {Reinhard}},
  \ and\ \bibinfo {author} {\bibfnamefont {E.}~\bibnamefont {Suraud}},\
  }\enquote {\bibinfo {title} {Similarities and differences between nuclei and
  metal clusters},}\ in\ \href {\doibase 10.1142/9789813209350_0001} {\emph
  {\bibinfo {booktitle} {Nuclear {{Particle Correlations}} and {{Cluster
  Physics}}}}}\ (\bibinfo  {publisher} {{World Scientific}},\ \bibinfo {year}
  {2017})\ pp.\ \bibinfo {pages} {3--27}\BibitemShut {NoStop}%
\bibitem [{\citenamefont {Manzhos}\ and\ \citenamefont
  {Carrington}(2021)}]{manzhos2021neural}%
  \BibitemOpen
  \bibfield  {author} {\bibinfo {author} {\bibfnamefont {S.}~\bibnamefont
  {Manzhos}}\ and\ \bibinfo {author} {\bibfnamefont {T.~J.}\ \bibnamefont
  {Carrington}},\ }\href {\doibase 10.1021/acs.chemrev.0c00665} {\bibfield
  {journal} {\bibinfo  {journal} {Chem. Rev.}\ }\textbf {\bibinfo {volume}
  {121}},\ \bibinfo {pages} {10187} (\bibinfo {year} {2021})}\BibitemShut
  {NoStop}%
\bibitem [{\citenamefont {Schlegel}(2003)}]{schlegel2003exploring}%
  \BibitemOpen
  \bibfield  {author} {\bibinfo {author} {\bibfnamefont {H.~B.}\ \bibnamefont
  {Schlegel}},\ }\href {\doibase 10.1002/jcc.10231} {\bibfield  {journal}
  {\bibinfo  {journal} {Journal of Computational Chemistry}\ }\textbf {\bibinfo
  {volume} {24}},\ \bibinfo {pages} {1514} (\bibinfo {year}
  {2003})}\BibitemShut {NoStop}%
\bibitem [{\citenamefont {Schunck}\ and\ \citenamefont
  {Regnier}(2022)}]{schunck2022theory}%
  \BibitemOpen
  \bibfield  {author} {\bibinfo {author} {\bibfnamefont {N.}~\bibnamefont
  {Schunck}}\ and\ \bibinfo {author} {\bibfnamefont {D.}~\bibnamefont
  {Regnier}},\ }\href {\doibase 10.1016/j.ppnp.2022.103963} {\bibfield
  {journal} {\bibinfo  {journal} {Progress in Particle and Nuclear Physics}\ ,\
  \bibinfo {pages} {103963}} (\bibinfo {year} {2022})}\BibitemShut {NoStop}%
\bibitem [{\citenamefont {Hollingsworth}\ and\ \citenamefont
  {Dror}(2018)}]{hollingsworth2018molecular}%
  \BibitemOpen
  \bibfield  {author} {\bibinfo {author} {\bibfnamefont {S.~A.}\ \bibnamefont
  {Hollingsworth}}\ and\ \bibinfo {author} {\bibfnamefont {R.~O.}\ \bibnamefont
  {Dror}},\ }\href {\doibase 10.1016/j.neuron.2018.08.011} {\bibfield
  {journal} {\bibinfo  {journal} {Neuron}\ }\textbf {\bibinfo {volume} {99}},\
  \bibinfo {pages} {1129} (\bibinfo {year} {2018})}\BibitemShut {NoStop}%
\bibitem [{\citenamefont {Verriere}\ and\ \citenamefont
  {Regnier}(2020)}]{verriere2020timedependent}%
  \BibitemOpen
  \bibfield  {author} {\bibinfo {author} {\bibfnamefont {M.}~\bibnamefont
  {Verriere}}\ and\ \bibinfo {author} {\bibfnamefont {D.}~\bibnamefont
  {Regnier}},\ }\href {\doibase 10.3389/fphy.2020.00233} {\bibfield  {journal}
  {\bibinfo  {journal} {Front. Phys.}\ }\textbf {\bibinfo {volume} {8}}
  (\bibinfo {year} {2020}),\ 10.3389/fphy.2020.00233}\BibitemShut {NoStop}%
\bibitem [{\citenamefont {Rogal}(2021)}]{rogal2021reaction}%
  \BibitemOpen
  \bibfield  {author} {\bibinfo {author} {\bibfnamefont {J.}~\bibnamefont
  {Rogal}},\ }\href {\doibase 10.1140/epjb/s10051-021-00233-5} {\bibfield
  {journal} {\bibinfo  {journal} {Eur. Phys. J. B}\ }\textbf {\bibinfo {volume}
  {94}},\ \bibinfo {pages} {223} (\bibinfo {year} {2021})}\BibitemShut
  {NoStop}%
\bibitem [{\citenamefont {Maeda}\ \emph {et~al.}(2015)\citenamefont {Maeda},
  \citenamefont {Harabuchi}, \citenamefont {Ono}, \citenamefont {Taketsugu},\
  and\ \citenamefont {Morokuma}}]{maeda2015intrinsic}%
  \BibitemOpen
  \bibfield  {author} {\bibinfo {author} {\bibfnamefont {S.}~\bibnamefont
  {Maeda}}, \bibinfo {author} {\bibfnamefont {Y.}~\bibnamefont {Harabuchi}},
  \bibinfo {author} {\bibfnamefont {Y.}~\bibnamefont {Ono}}, \bibinfo {author}
  {\bibfnamefont {T.}~\bibnamefont {Taketsugu}}, \ and\ \bibinfo {author}
  {\bibfnamefont {K.}~\bibnamefont {Morokuma}},\ }\href {\doibase
  10.1002/qua.24757} {\bibfield  {journal} {\bibinfo  {journal} {International
  Journal of Quantum Chemistry}\ }\textbf {\bibinfo {volume} {115}},\ \bibinfo
  {pages} {258} (\bibinfo {year} {2015})}\BibitemShut {NoStop}%
\bibitem [{\citenamefont {Stewart}\ \emph {et~al.}(1987)\citenamefont
  {Stewart}, \citenamefont {Davis},\ and\ \citenamefont
  {Burggraf}}]{stewart1987semiempirical}%
  \BibitemOpen
  \bibfield  {author} {\bibinfo {author} {\bibfnamefont {J.~J.}\ \bibnamefont
  {Stewart}}, \bibinfo {author} {\bibfnamefont {L.~P.}\ \bibnamefont {Davis}},
  \ and\ \bibinfo {author} {\bibfnamefont {L.~W.}\ \bibnamefont {Burggraf}},\
  }\href {\doibase 10.1002/jcc.540080808} {\bibfield  {journal} {\bibinfo
  {journal} {Journal of Computational Chemistry}\ }\textbf {\bibinfo {volume}
  {8}},\ \bibinfo {pages} {1117} (\bibinfo {year} {1987})}\BibitemShut
  {NoStop}%
\bibitem [{\citenamefont {Denzel}\ \emph {et~al.}(2019)\citenamefont {Denzel},
  \citenamefont {Haasdonk},\ and\ \citenamefont
  {K{\"a}stner}}]{denzel2019gaussian}%
  \BibitemOpen
  \bibfield  {author} {\bibinfo {author} {\bibfnamefont {A.}~\bibnamefont
  {Denzel}}, \bibinfo {author} {\bibfnamefont {B.}~\bibnamefont {Haasdonk}}, \
  and\ \bibinfo {author} {\bibfnamefont {J.}~\bibnamefont {K{\"a}stner}},\
  }\href {\doibase 10.1021/acs.jpca.9b08239} {\bibfield  {journal} {\bibinfo
  {journal} {J. Phys. Chem. A}\ }\textbf {\bibinfo {volume} {123}},\ \bibinfo
  {pages} {9600} (\bibinfo {year} {2019})}\BibitemShut {NoStop}%
\bibitem [{\citenamefont {Tiwary}\ and\ \citenamefont
  {Parrinello}(2013)}]{tiwary2013metadynamics}%
  \BibitemOpen
  \bibfield  {author} {\bibinfo {author} {\bibfnamefont {P.}~\bibnamefont
  {Tiwary}}\ and\ \bibinfo {author} {\bibfnamefont {M.}~\bibnamefont
  {Parrinello}},\ }\href {\doibase 10.1103/PhysRevLett.111.230602} {\bibfield
  {journal} {\bibinfo  {journal} {Phys. Rev. Lett.}\ }\textbf {\bibinfo
  {volume} {111}},\ \bibinfo {pages} {230602} (\bibinfo {year}
  {2013})}\BibitemShut {NoStop}%
\bibitem [{\citenamefont {Bussi}\ and\ \citenamefont
  {Laio}(2020)}]{bussi2020using}%
  \BibitemOpen
  \bibfield  {author} {\bibinfo {author} {\bibfnamefont {G.}~\bibnamefont
  {Bussi}}\ and\ \bibinfo {author} {\bibfnamefont {A.}~\bibnamefont {Laio}},\
  }\href {\doibase 10.1038/s42254-020-0153-0} {\bibfield  {journal} {\bibinfo
  {journal} {Nat. Rev. Phys.}\ }\textbf {\bibinfo {volume} {2}},\ \bibinfo
  {pages} {200} (\bibinfo {year} {2020})}\BibitemShut {NoStop}%
\bibitem [{\citenamefont {Evenhuis}\ and\ \citenamefont
  {Mart{\'i}nez}(2011)}]{evenhuis2011scheme}%
  \BibitemOpen
  \bibfield  {author} {\bibinfo {author} {\bibfnamefont {C.}~\bibnamefont
  {Evenhuis}}\ and\ \bibinfo {author} {\bibfnamefont {T.~J.}\ \bibnamefont
  {Mart{\'i}nez}},\ }\href {\doibase 10.1063/1.3660686} {\bibfield  {journal}
  {\bibinfo  {journal} {J. Chem. Phys.}\ }\textbf {\bibinfo {volume} {135}},\
  \bibinfo {pages} {224110} (\bibinfo {year} {2011})}\BibitemShut {NoStop}%
\bibitem [{\citenamefont {Bender}(2008)}]{bender2008going}%
  \BibitemOpen
  \bibfield  {author} {\bibinfo {author} {\bibfnamefont {M.}~\bibnamefont
  {Bender}},\ }\href {\doibase 10.1140/epjst/e2008-00619-9} {\bibfield
  {journal} {\bibinfo  {journal} {The European Physical Journal Special
  Topics}\ }\textbf {\bibinfo {volume} {156}},\ \bibinfo {pages} {217}
  (\bibinfo {year} {2008})}\BibitemShut {NoStop}%
\bibitem [{\citenamefont {Girod}\ and\ \citenamefont
  {Schuck}(2013)}]{girod2013alphaparticle}%
  \BibitemOpen
  \bibfield  {author} {\bibinfo {author} {\bibfnamefont {M.}~\bibnamefont
  {Girod}}\ and\ \bibinfo {author} {\bibfnamefont {P.}~\bibnamefont {Schuck}},\
  }\href {\doibase 10.1103/PhysRevLett.111.132503} {\bibfield  {journal}
  {\bibinfo  {journal} {Phys. Rev. Lett.}\ }\textbf {\bibinfo {volume} {111}},\
  \bibinfo {pages} {132503} (\bibinfo {year} {2013})}\BibitemShut {NoStop}%
\bibitem [{\citenamefont {Ebran}\ \emph {et~al.}(2020)\citenamefont {Ebran},
  \citenamefont {Girod}, \citenamefont {Khan}, \citenamefont {Lasseri},\ and\
  \citenamefont {Schuck}}]{ebran2020ensurematha}%
  \BibitemOpen
  \bibfield  {author} {\bibinfo {author} {\bibfnamefont {J.-P.}\ \bibnamefont
  {Ebran}}, \bibinfo {author} {\bibfnamefont {M.}~\bibnamefont {Girod}},
  \bibinfo {author} {\bibfnamefont {E.}~\bibnamefont {Khan}}, \bibinfo {author}
  {\bibfnamefont {R.~D.}\ \bibnamefont {Lasseri}}, \ and\ \bibinfo {author}
  {\bibfnamefont {P.}~\bibnamefont {Schuck}},\ }\href {\doibase
  10.1103/PhysRevC.102.014305} {\bibfield  {journal} {\bibinfo  {journal}
  {Phys. Rev. C}\ }\textbf {\bibinfo {volume} {102}},\ \bibinfo {pages}
  {014305} (\bibinfo {year} {2020})}\BibitemShut {NoStop}%
\bibitem [{\citenamefont {Dubray}\ and\ \citenamefont
  {Regnier}(2012)}]{dubray2012numerical}%
  \BibitemOpen
  \bibfield  {author} {\bibinfo {author} {\bibfnamefont {N.}~\bibnamefont
  {Dubray}}\ and\ \bibinfo {author} {\bibfnamefont {D.}~\bibnamefont
  {Regnier}},\ }\href {\doibase 10.1016/j.cpc.2012.05.001} {\bibfield
  {journal} {\bibinfo  {journal} {Computer Physics Communications}\ }\textbf
  {\bibinfo {volume} {183}},\ \bibinfo {pages} {2035} (\bibinfo {year}
  {2012})}\BibitemShut {NoStop}%
\bibitem [{\citenamefont {Regnier}\ \emph {et~al.}(2019)\citenamefont
  {Regnier}, \citenamefont {Dubray},\ and\ \citenamefont
  {Schunck}}]{regnier2019asymmetric}%
  \BibitemOpen
  \bibfield  {author} {\bibinfo {author} {\bibfnamefont {D.}~\bibnamefont
  {Regnier}}, \bibinfo {author} {\bibfnamefont {N.}~\bibnamefont {Dubray}}, \
  and\ \bibinfo {author} {\bibfnamefont {N.}~\bibnamefont {Schunck}},\ }\href
  {\doibase 10.1103/PhysRevC.99.024611} {\bibfield  {journal} {\bibinfo
  {journal} {Phys. Rev. C}\ }\textbf {\bibinfo {volume} {99}},\ \bibinfo
  {pages} {024611} (\bibinfo {year} {2019})}\BibitemShut {NoStop}%
\bibitem [{\citenamefont {Ho}\ and\ \citenamefont
  {Rabitz}(1996)}]{ho1996general}%
  \BibitemOpen
  \bibfield  {author} {\bibinfo {author} {\bibfnamefont {T.-S.}\ \bibnamefont
  {Ho}}\ and\ \bibinfo {author} {\bibfnamefont {H.}~\bibnamefont {Rabitz}},\
  }\href {\doibase 10.1063/1.470984} {\bibfield  {journal} {\bibinfo  {journal}
  {J. Chem. Phys.}\ }\textbf {\bibinfo {volume} {104}},\ \bibinfo {pages}
  {2584} (\bibinfo {year} {1996})}\BibitemShut {NoStop}%
\bibitem [{\citenamefont {Hurd}\ \emph {et~al.}(2010)\citenamefont {Hurd},
  \citenamefont {Cusati},\ and\ \citenamefont {Persico}}]{hurd2010trajectory}%
  \BibitemOpen
  \bibfield  {author} {\bibinfo {author} {\bibfnamefont {P.}~\bibnamefont
  {Hurd}}, \bibinfo {author} {\bibfnamefont {T.}~\bibnamefont {Cusati}}, \ and\
  \bibinfo {author} {\bibfnamefont {M.}~\bibnamefont {Persico}},\ }\href
  {\doibase 10.1016/j.jcp.2009.11.025} {\bibfield  {journal} {\bibinfo
  {journal} {Journal of Computational Physics}\ }\textbf {\bibinfo {volume}
  {229}},\ \bibinfo {pages} {2109} (\bibinfo {year} {2010})}\BibitemShut
  {NoStop}%
\bibitem [{\citenamefont {Galv{\~a}o}\ \emph {et~al.}(2016)\citenamefont
  {Galv{\~a}o}, \citenamefont {Mota},\ and\ \citenamefont
  {Varandas}}]{galvao2016modeling}%
  \BibitemOpen
  \bibfield  {author} {\bibinfo {author} {\bibfnamefont {B.~R.~L.}\
  \bibnamefont {Galv{\~a}o}}, \bibinfo {author} {\bibfnamefont {V.~C.}\
  \bibnamefont {Mota}}, \ and\ \bibinfo {author} {\bibfnamefont {A.~J.~C.}\
  \bibnamefont {Varandas}},\ }\href {\doibase 10.1016/j.cplett.2016.07.029}
  {\bibfield  {journal} {\bibinfo  {journal} {Chemical Physics Letters}\
  }\textbf {\bibinfo {volume} {660}},\ \bibinfo {pages} {55} (\bibinfo {year}
  {2016})}\BibitemShut {NoStop}%
\bibitem [{\citenamefont {Curchod}\ and\ \citenamefont
  {Mart{\'i}nez}(2018)}]{curchod2018initio}%
  \BibitemOpen
  \bibfield  {author} {\bibinfo {author} {\bibfnamefont {B.~F.~E.}\
  \bibnamefont {Curchod}}\ and\ \bibinfo {author} {\bibfnamefont {T.~J.}\
  \bibnamefont {Mart{\'i}nez}},\ }\href {\doibase 10.1021/acs.chemrev.7b00423}
  {\bibfield  {journal} {\bibinfo  {journal} {Chem. Rev.}\ }\textbf {\bibinfo
  {volume} {118}},\ \bibinfo {pages} {3305} (\bibinfo {year}
  {2018})}\BibitemShut {NoStop}%
\bibitem [{\citenamefont {Varandas}(2009)}]{varandas2009simple}%
  \BibitemOpen
  \bibfield  {author} {\bibinfo {author} {\bibfnamefont {A.~J.~C.}\
  \bibnamefont {Varandas}},\ }\href {\doibase 10.1016/j.cplett.2009.02.028}
  {\bibfield  {journal} {\bibinfo  {journal} {Chemical Physics Letters}\
  }\textbf {\bibinfo {volume} {471}},\ \bibinfo {pages} {315} (\bibinfo {year}
  {2009})}\BibitemShut {NoStop}%
\bibitem [{\citenamefont {Bernard}\ \emph {et~al.}(2011)\citenamefont
  {Bernard}, \citenamefont {Goutte}, \citenamefont {Gogny},\ and\ \citenamefont
  {Younes}}]{bernard2011microscopic}%
  \BibitemOpen
  \bibfield  {author} {\bibinfo {author} {\bibfnamefont {R.}~\bibnamefont
  {Bernard}}, \bibinfo {author} {\bibfnamefont {H.}~\bibnamefont {Goutte}},
  \bibinfo {author} {\bibfnamefont {D.}~\bibnamefont {Gogny}}, \ and\ \bibinfo
  {author} {\bibfnamefont {W.}~\bibnamefont {Younes}},\ }\href {\doibase
  10.1103/PhysRevC.84.044308} {\bibfield  {journal} {\bibinfo  {journal} {Phys.
  Rev. C}\ }\textbf {\bibinfo {volume} {84}},\ \bibinfo {pages} {044308}
  (\bibinfo {year} {2011})}\BibitemShut {NoStop}%
\bibitem [{\citenamefont {Dietrich}\ \emph {et~al.}(2010)\citenamefont
  {Dietrich}, \citenamefont {Niez},\ and\ \citenamefont
  {Berger}}]{dietrich2010microscopic}%
  \BibitemOpen
  \bibfield  {author} {\bibinfo {author} {\bibfnamefont {K.}~\bibnamefont
  {Dietrich}}, \bibinfo {author} {\bibfnamefont {J.-J.}\ \bibnamefont {Niez}},
  \ and\ \bibinfo {author} {\bibfnamefont {J.-F.}\ \bibnamefont {Berger}},\
  }\href {\doibase 10.1142/S0218301310014935} {\bibfield  {journal} {\bibinfo
  {journal} {Int. J. Mod. Phys. E}\ }\textbf {\bibinfo {volume} {19}},\
  \bibinfo {pages} {521} (\bibinfo {year} {2010})}\BibitemShut {NoStop}%
\bibitem [{\citenamefont {Zhao}\ \emph {et~al.}(2019)\citenamefont {Zhao},
  \citenamefont {Nik{\v s}i{\'c}}, \citenamefont {Vretenar},\ and\
  \citenamefont {Zhou}}]{zhao2019microscopic}%
  \BibitemOpen
  \bibfield  {author} {\bibinfo {author} {\bibfnamefont {J.}~\bibnamefont
  {Zhao}}, \bibinfo {author} {\bibfnamefont {T.}~\bibnamefont {Nik{\v
  s}i{\'c}}}, \bibinfo {author} {\bibfnamefont {D.}~\bibnamefont {Vretenar}}, \
  and\ \bibinfo {author} {\bibfnamefont {S.-G.}\ \bibnamefont {Zhou}},\ }\href
  {\doibase 10.1103/PhysRevC.99.014618} {\bibfield  {journal} {\bibinfo
  {journal} {Phys. Rev. C}\ }\textbf {\bibinfo {volume} {99}},\ \bibinfo
  {pages} {014618} (\bibinfo {year} {2019})}\BibitemShut {NoStop}%
\bibitem [{\citenamefont {Lau}\ \emph {et~al.}(2022)\citenamefont {Lau},
  \citenamefont {Bernard},\ and\ \citenamefont {Simenel}}]{lau2022smoothing}%
  \BibitemOpen
  \bibfield  {author} {\bibinfo {author} {\bibfnamefont {N.-W.~T.}\
  \bibnamefont {Lau}}, \bibinfo {author} {\bibfnamefont {R.~N.}\ \bibnamefont
  {Bernard}}, \ and\ \bibinfo {author} {\bibfnamefont {C.}~\bibnamefont
  {Simenel}},\ }\href {\doibase 10.1103/PhysRevC.105.034617} {\bibfield
  {journal} {\bibinfo  {journal} {Phys. Rev. C}\ }\textbf {\bibinfo {volume}
  {105}},\ \bibinfo {pages} {034617} (\bibinfo {year} {2022})}\BibitemShut
  {NoStop}%
\bibitem [{\citenamefont {Ma}\ and\ \citenamefont
  {Dinner}(2005)}]{ma2005automatic}%
  \BibitemOpen
  \bibfield  {author} {\bibinfo {author} {\bibfnamefont {A.}~\bibnamefont
  {Ma}}\ and\ \bibinfo {author} {\bibfnamefont {A.~R.}\ \bibnamefont
  {Dinner}},\ }\href {\doibase 10.1021/jp045546c} {\bibfield  {journal}
  {\bibinfo  {journal} {J. Phys. Chem. B}\ }\textbf {\bibinfo {volume} {109}},\
  \bibinfo {pages} {6769} (\bibinfo {year} {2005})}\BibitemShut {NoStop}%
\bibitem [{\citenamefont {Baima}\ \emph {et~al.}(2022)\citenamefont {Baima},
  \citenamefont {Goryaeva}, \citenamefont {Swinburne}, \citenamefont {Maillet},
  \citenamefont {Nastar},\ and\ \citenamefont
  {Marinica}}]{baima2022capabilities}%
  \BibitemOpen
  \bibfield  {author} {\bibinfo {author} {\bibfnamefont {J.}~\bibnamefont
  {Baima}}, \bibinfo {author} {\bibfnamefont {A.~M.}\ \bibnamefont {Goryaeva}},
  \bibinfo {author} {\bibfnamefont {T.~D.}\ \bibnamefont {Swinburne}}, \bibinfo
  {author} {\bibfnamefont {J.-B.}\ \bibnamefont {Maillet}}, \bibinfo {author}
  {\bibfnamefont {M.}~\bibnamefont {Nastar}}, \ and\ \bibinfo {author}
  {\bibfnamefont {M.-C.}\ \bibnamefont {Marinica}},\ }\href {\doibase
  10.1039/D2CP01917E} {\bibfield  {journal} {\bibinfo  {journal} {Phys. Chem.
  Chem. Phys.}\ }\textbf {\bibinfo {volume} {24}},\ \bibinfo {pages} {23152}
  (\bibinfo {year} {2022})}\BibitemShut {NoStop}%
\bibitem [{\citenamefont {Belkacemi}\ \emph {et~al.}(2022)\citenamefont
  {Belkacemi}, \citenamefont {Gkeka}, \citenamefont {Leli{\`e}vre},\ and\
  \citenamefont {Stoltz}}]{belkacemi2022chasing}%
  \BibitemOpen
  \bibfield  {author} {\bibinfo {author} {\bibfnamefont {Z.}~\bibnamefont
  {Belkacemi}}, \bibinfo {author} {\bibfnamefont {P.}~\bibnamefont {Gkeka}},
  \bibinfo {author} {\bibfnamefont {T.}~\bibnamefont {Leli{\`e}vre}}, \ and\
  \bibinfo {author} {\bibfnamefont {G.}~\bibnamefont {Stoltz}},\ }\href
  {\doibase 10.1021/acs.jctc.1c00415} {\bibfield  {journal} {\bibinfo
  {journal} {J. Chem. Theory Comput.}\ }\textbf {\bibinfo {volume} {18}},\
  \bibinfo {pages} {59} (\bibinfo {year} {2022})}\BibitemShut {NoStop}%
\bibitem [{\citenamefont {Verriere}\ \emph {et~al.}(2022)\citenamefont
  {Verriere}, \citenamefont {Schunck}, \citenamefont {Kim}, \citenamefont
  {Marevi{\'c}}, \citenamefont {Quinlan}, \citenamefont {Ngo}, \citenamefont
  {Regnier},\ and\ \citenamefont {Lasseri}}]{verriere2022building}%
  \BibitemOpen
  \bibfield  {author} {\bibinfo {author} {\bibfnamefont {M.}~\bibnamefont
  {Verriere}}, \bibinfo {author} {\bibfnamefont {N.}~\bibnamefont {Schunck}},
  \bibinfo {author} {\bibfnamefont {I.}~\bibnamefont {Kim}}, \bibinfo {author}
  {\bibfnamefont {P.}~\bibnamefont {Marevi{\'c}}}, \bibinfo {author}
  {\bibfnamefont {K.}~\bibnamefont {Quinlan}}, \bibinfo {author} {\bibfnamefont
  {M.~N.}\ \bibnamefont {Ngo}}, \bibinfo {author} {\bibfnamefont
  {D.}~\bibnamefont {Regnier}}, \ and\ \bibinfo {author} {\bibfnamefont
  {R.~D.}\ \bibnamefont {Lasseri}},\ }\href@noop {} {\bibfield  {journal}
  {\bibinfo  {journal} {Frontiers in Physics}\ }\textbf {\bibinfo {volume}
  {10}} (\bibinfo {year} {2022})}\BibitemShut {NoStop}%
\bibitem [{\citenamefont {Carleo}\ \emph {et~al.}(2018)\citenamefont {Carleo},
  \citenamefont {Nomura},\ and\ \citenamefont
  {Imada}}]{carleo2018constructing}%
  \BibitemOpen
  \bibfield  {author} {\bibinfo {author} {\bibfnamefont {G.}~\bibnamefont
  {Carleo}}, \bibinfo {author} {\bibfnamefont {Y.}~\bibnamefont {Nomura}}, \
  and\ \bibinfo {author} {\bibfnamefont {M.}~\bibnamefont {Imada}},\ }\href
  {\doibase 10.1038/s41467-018-07520-3} {\bibfield  {journal} {\bibinfo
  {journal} {Nat Commun}\ }\textbf {\bibinfo {volume} {9}},\ \bibinfo {pages}
  {1} (\bibinfo {year} {2018})}\BibitemShut {NoStop}%
\bibitem [{\citenamefont {Carleo}\ \emph {et~al.}(2019)\citenamefont {Carleo},
  \citenamefont {Cirac}, \citenamefont {Cranmer}, \citenamefont {Daudet},
  \citenamefont {Schuld}, \citenamefont {Tishby}, \citenamefont
  {{Vogt-Maranto}},\ and\ \citenamefont {Zdeborov{\'a}}}]{carleo2019machine}%
  \BibitemOpen
  \bibfield  {author} {\bibinfo {author} {\bibfnamefont {G.}~\bibnamefont
  {Carleo}}, \bibinfo {author} {\bibfnamefont {I.}~\bibnamefont {Cirac}},
  \bibinfo {author} {\bibfnamefont {K.}~\bibnamefont {Cranmer}}, \bibinfo
  {author} {\bibfnamefont {L.}~\bibnamefont {Daudet}}, \bibinfo {author}
  {\bibfnamefont {M.}~\bibnamefont {Schuld}}, \bibinfo {author} {\bibfnamefont
  {N.}~\bibnamefont {Tishby}}, \bibinfo {author} {\bibfnamefont
  {L.}~\bibnamefont {{Vogt-Maranto}}}, \ and\ \bibinfo {author} {\bibfnamefont
  {L.}~\bibnamefont {Zdeborov{\'a}}},\ }\href {\doibase
  10.1103/RevModPhys.91.045002} {\bibfield  {journal} {\bibinfo  {journal}
  {Rev. Mod. Phys.}\ }\textbf {\bibinfo {volume} {91}},\ \bibinfo {pages}
  {045002} (\bibinfo {year} {2019})}\BibitemShut {NoStop}%
\bibitem [{\citenamefont {Sch{\"u}tt}\ \emph {et~al.}(2019)\citenamefont
  {Sch{\"u}tt}, \citenamefont {Gastegger}, \citenamefont {Tkatchenko},
  \citenamefont {M{\"u}ller},\ and\ \citenamefont
  {Maurer}}]{schutt2019unifying}%
  \BibitemOpen
  \bibfield  {author} {\bibinfo {author} {\bibfnamefont {K.~T.}\ \bibnamefont
  {Sch{\"u}tt}}, \bibinfo {author} {\bibfnamefont {M.}~\bibnamefont
  {Gastegger}}, \bibinfo {author} {\bibfnamefont {A.}~\bibnamefont
  {Tkatchenko}}, \bibinfo {author} {\bibfnamefont {K.-R.}\ \bibnamefont
  {M{\"u}ller}}, \ and\ \bibinfo {author} {\bibfnamefont {R.~J.}\ \bibnamefont
  {Maurer}},\ }\href {\doibase 10.1038/s41467-019-12875-2} {\bibfield
  {journal} {\bibinfo  {journal} {Nat Commun}\ }\textbf {\bibinfo {volume}
  {10}},\ \bibinfo {pages} {5024} (\bibinfo {year} {2019})}\BibitemShut
  {NoStop}%
\bibitem [{\citenamefont {Companys~Franzke}\ \emph {et~al.}(2022)\citenamefont
  {Companys~Franzke}, \citenamefont {Tichai}, \citenamefont {Hebeler},\ and\
  \citenamefont {Schwenk}}]{companysfranzke2022excited}%
  \BibitemOpen
  \bibfield  {author} {\bibinfo {author} {\bibfnamefont {M.}~\bibnamefont
  {Companys~Franzke}}, \bibinfo {author} {\bibfnamefont {A.}~\bibnamefont
  {Tichai}}, \bibinfo {author} {\bibfnamefont {K.}~\bibnamefont {Hebeler}}, \
  and\ \bibinfo {author} {\bibfnamefont {A.}~\bibnamefont {Schwenk}},\ }\href
  {\doibase 10.1016/j.physletb.2022.137101} {\bibfield  {journal} {\bibinfo
  {journal} {Physics Letters B}\ }\textbf {\bibinfo {volume} {830}},\ \bibinfo
  {pages} {137101} (\bibinfo {year} {2022})}\BibitemShut {NoStop}%
\bibitem [{\citenamefont {Cassella}\ \emph {et~al.}(2023)\citenamefont
  {Cassella}, \citenamefont {Sutterud}, \citenamefont {Azadi}, \citenamefont
  {Drummond}, \citenamefont {Pfau}, \citenamefont {Spencer},\ and\
  \citenamefont {Foulkes}}]{cassella2023discovering}%
  \BibitemOpen
  \bibfield  {author} {\bibinfo {author} {\bibfnamefont {G.}~\bibnamefont
  {Cassella}}, \bibinfo {author} {\bibfnamefont {H.}~\bibnamefont {Sutterud}},
  \bibinfo {author} {\bibfnamefont {S.}~\bibnamefont {Azadi}}, \bibinfo
  {author} {\bibfnamefont {N.~D.}\ \bibnamefont {Drummond}}, \bibinfo {author}
  {\bibfnamefont {D.}~\bibnamefont {Pfau}}, \bibinfo {author} {\bibfnamefont
  {J.~S.}\ \bibnamefont {Spencer}}, \ and\ \bibinfo {author} {\bibfnamefont
  {W.~M.~C.}\ \bibnamefont {Foulkes}},\ }\href {\doibase
  10.1103/PhysRevLett.130.036401} {\bibfield  {journal} {\bibinfo  {journal}
  {Phys. Rev. Lett.}\ }\textbf {\bibinfo {volume} {130}},\ \bibinfo {pages}
  {036401} (\bibinfo {year} {2023})}\BibitemShut {NoStop}%
\bibitem [{\citenamefont {Asif}\ \emph {et~al.}(2021)\citenamefont {Asif},
  \citenamefont {Sarker}, \citenamefont {Chakrabortty}, \citenamefont {Ryan},
  \citenamefont {Ahamed}, \citenamefont {Saha}, \citenamefont {Badal},
  \citenamefont {Das}, \citenamefont {Ali}, \citenamefont {Moyeen},
  \citenamefont {Islam},\ and\ \citenamefont {Tasneem}}]{asif2021graph}%
  \BibitemOpen
  \bibfield  {author} {\bibinfo {author} {\bibfnamefont {N.~A.}\ \bibnamefont
  {Asif}}, \bibinfo {author} {\bibfnamefont {Y.}~\bibnamefont {Sarker}},
  \bibinfo {author} {\bibfnamefont {R.~K.}\ \bibnamefont {Chakrabortty}},
  \bibinfo {author} {\bibfnamefont {M.~J.}\ \bibnamefont {Ryan}}, \bibinfo
  {author} {\bibfnamefont {M.~H.}\ \bibnamefont {Ahamed}}, \bibinfo {author}
  {\bibfnamefont {D.~K.}\ \bibnamefont {Saha}}, \bibinfo {author}
  {\bibfnamefont {F.~R.}\ \bibnamefont {Badal}}, \bibinfo {author}
  {\bibfnamefont {S.~K.}\ \bibnamefont {Das}}, \bibinfo {author} {\bibfnamefont
  {M.~F.}\ \bibnamefont {Ali}}, \bibinfo {author} {\bibfnamefont {S.~I.}\
  \bibnamefont {Moyeen}}, \bibinfo {author} {\bibfnamefont {M.~R.}\
  \bibnamefont {Islam}}, \ and\ \bibinfo {author} {\bibfnamefont
  {Z.}~\bibnamefont {Tasneem}},\ }\href {\doibase 10.1109/ACCESS.2021.3071274}
  {\bibfield  {journal} {\bibinfo  {journal} {IEEE Access}\ }\textbf {\bibinfo
  {volume} {9}},\ \bibinfo {pages} {60588} (\bibinfo {year}
  {2021})}\BibitemShut {NoStop}%
\bibitem [{\citenamefont {Zhang}\ \emph {et~al.}(2018)\citenamefont {Zhang},
  \citenamefont {Zhu}, \citenamefont {Heath~Jr.},\ and\ \citenamefont
  {Huang}}]{zhang2018grassmannian}%
  \BibitemOpen
  \bibfield  {author} {\bibinfo {author} {\bibfnamefont {J.}~\bibnamefont
  {Zhang}}, \bibinfo {author} {\bibfnamefont {G.}~\bibnamefont {Zhu}}, \bibinfo
  {author} {\bibfnamefont {R.~W.}\ \bibnamefont {Heath~Jr.}}, \ and\ \bibinfo
  {author} {\bibfnamefont {K.}~\bibnamefont {Huang}},\ }\href@noop {}
  {\bibfield  {journal} {\bibinfo  {journal} {arXiv:1808.02229 [cs, eess, math,
  stat]}\ } (\bibinfo {year} {2018})}\BibitemShut {NoStop}%
\bibitem [{\citenamefont {Chakraborty}\ \emph {et~al.}(2022)\citenamefont
  {Chakraborty}, \citenamefont {Bouza}, \citenamefont {Manton},\ and\
  \citenamefont {Vemuri}}]{chakraborty2022manifoldnet}%
  \BibitemOpen
  \bibfield  {author} {\bibinfo {author} {\bibfnamefont {R.}~\bibnamefont
  {Chakraborty}}, \bibinfo {author} {\bibfnamefont {J.}~\bibnamefont {Bouza}},
  \bibinfo {author} {\bibfnamefont {J.~H.}\ \bibnamefont {Manton}}, \ and\
  \bibinfo {author} {\bibfnamefont {B.~C.}\ \bibnamefont {Vemuri}},\ }\href
  {\doibase 10.1109/TPAMI.2020.3003846} {\bibfield  {journal} {\bibinfo
  {journal} {IEEE Transactions on Pattern Analysis and Machine Intelligence}\
  }\textbf {\bibinfo {volume} {44}},\ \bibinfo {pages} {799} (\bibinfo {year}
  {2022})}\BibitemShut {NoStop}%
\bibitem [{\citenamefont {Machleidt}\ and\ \citenamefont
  {Sammarruca}(2016)}]{machleidt2016chiral}%
  \BibitemOpen
  \bibfield  {author} {\bibinfo {author} {\bibfnamefont {R.}~\bibnamefont
  {Machleidt}}\ and\ \bibinfo {author} {\bibfnamefont {F.}~\bibnamefont
  {Sammarruca}},\ }\href {\doibase 10.1088/0031-8949/91/8/083007} {\bibfield
  {journal} {\bibinfo  {journal} {Phys. Scr.}\ }\textbf {\bibinfo {volume}
  {91}},\ \bibinfo {pages} {083007} (\bibinfo {year} {2016})}\BibitemShut
  {NoStop}%
\bibitem [{\citenamefont {Bender}\ \emph {et~al.}(2003)\citenamefont {Bender},
  \citenamefont {Heenen},\ and\ \citenamefont
  {Reinhard}}]{bender2003selfconsistent}%
  \BibitemOpen
  \bibfield  {author} {\bibinfo {author} {\bibfnamefont {M.}~\bibnamefont
  {Bender}}, \bibinfo {author} {\bibfnamefont {P.~H.}\ \bibnamefont {Heenen}},
  \ and\ \bibinfo {author} {\bibfnamefont {P.~G.}\ \bibnamefont {Reinhard}},\
  }\href@noop {} {\bibfield  {journal} {\bibinfo  {journal} {Rev. Mod. Phys.}\
  }\textbf {\bibinfo {volume} {75}},\ \bibinfo {pages} {121} (\bibinfo {year}
  {2003})}\BibitemShut {NoStop}%
\bibitem [{\citenamefont {Carollo}\ \emph {et~al.}(2020)\citenamefont
  {Carollo}, \citenamefont {Valenti},\ and\ \citenamefont
  {Spagnolo}}]{carollo2020geometry}%
  \BibitemOpen
  \bibfield  {author} {\bibinfo {author} {\bibfnamefont {A.}~\bibnamefont
  {Carollo}}, \bibinfo {author} {\bibfnamefont {D.}~\bibnamefont {Valenti}}, \
  and\ \bibinfo {author} {\bibfnamefont {B.}~\bibnamefont {Spagnolo}},\ }\href
  {\doibase 10.1016/j.physrep.2019.11.002} {\bibfield  {journal} {\bibinfo
  {journal} {Physics Reports}\ }\bibinfo {series} {Geometry of Quantum Phase
  Transitions},\ \textbf {\bibinfo {volume} {838}},\ \bibinfo {pages} {1}
  (\bibinfo {year} {2020})}\BibitemShut {NoStop}%
\bibitem [{\citenamefont {Sadhukhan}(2020)}]{sadhukhan2020microscopic}%
  \BibitemOpen
  \bibfield  {author} {\bibinfo {author} {\bibfnamefont {J.}~\bibnamefont
  {Sadhukhan}},\ }\href {\doibase 10.3389/fphy.2020.567171} {\bibfield
  {journal} {\bibinfo  {journal} {Front. Phys.}\ }\textbf {\bibinfo {volume}
  {8}},\ \bibinfo {pages} {567171} (\bibinfo {year} {2020})}\BibitemShut
  {NoStop}%
\bibitem [{\citenamefont {Aoto}\ and\ \citenamefont {{da
  Silva}}(2020)}]{aoto2020calculating}%
  \BibitemOpen
  \bibfield  {author} {\bibinfo {author} {\bibfnamefont {Y.~A.}\ \bibnamefont
  {Aoto}}\ and\ \bibinfo {author} {\bibfnamefont {M.~F.}\ \bibnamefont {{da
  Silva}}},\ }\href {\doibase 10.1103/PhysRevA.102.052803} {\bibfield
  {journal} {\bibinfo  {journal} {Phys. Rev. A}\ }\textbf {\bibinfo {volume}
  {102}},\ \bibinfo {pages} {052803} (\bibinfo {year} {2020})}\BibitemShut
  {NoStop}%
\bibitem [{\citenamefont {H{\"u}per}\ \emph {et~al.}(2013)\citenamefont
  {H{\"u}per}, \citenamefont {Kleinsteuber},\ and\ \citenamefont
  {Shen}}]{huper2013averaginga}%
  \BibitemOpen
  \bibfield  {author} {\bibinfo {author} {\bibfnamefont {K.}~\bibnamefont
  {H{\"u}per}}, \bibinfo {author} {\bibfnamefont {M.}~\bibnamefont
  {Kleinsteuber}}, \ and\ \bibinfo {author} {\bibfnamefont {H.}~\bibnamefont
  {Shen}},\ }\href {\doibase 10.1016/j.sigpro.2012.08.020} {\bibfield
  {journal} {\bibinfo  {journal} {Signal Processing}\ }\textbf {\bibinfo
  {volume} {93}},\ \bibinfo {pages} {459} (\bibinfo {year} {2013})}\BibitemShut
  {NoStop}%
\bibitem [{\citenamefont {Townsend}\ \emph {et~al.}(2016)\citenamefont
  {Townsend}, \citenamefont {Koep},\ and\ \citenamefont
  {Weichwald}}]{JMLR:v17:16-177}%
  \BibitemOpen
  \bibfield  {author} {\bibinfo {author} {\bibfnamefont {J.}~\bibnamefont
  {Townsend}}, \bibinfo {author} {\bibfnamefont {N.}~\bibnamefont {Koep}}, \
  and\ \bibinfo {author} {\bibfnamefont {S.}~\bibnamefont {Weichwald}},\ }\href
  {http://jmlr.org/papers/v17/16-177.html} {\bibfield  {journal} {\bibinfo
  {journal} {Journal of Machine Learning Research}\ }\textbf {\bibinfo {volume}
  {17}},\ \bibinfo {pages} {1–5} (\bibinfo {year} {2016})}\BibitemShut
  {NoStop}%
\bibitem [{\citenamefont {Kingma}\ and\ \citenamefont
  {Welling}(2019)}]{kingma2019introduction}%
  \BibitemOpen
  \bibfield  {author} {\bibinfo {author} {\bibfnamefont {D.~P.}\ \bibnamefont
  {Kingma}}\ and\ \bibinfo {author} {\bibfnamefont {M.}~\bibnamefont
  {Welling}},\ }\href {\doibase 10.1561/2200000056} {\bibfield  {journal}
  {\bibinfo  {journal} {MAL}\ }\textbf {\bibinfo {volume} {12}},\ \bibinfo
  {pages} {307} (\bibinfo {year} {2019})}\BibitemShut {NoStop}%
\bibitem [{\citenamefont {Snoek}\ \emph {et~al.}(2012)\citenamefont {Snoek},
  \citenamefont {Larochelle},\ and\ \citenamefont {Adams}}]{NIPS2012_05311655}%
  \BibitemOpen
  \bibfield  {author} {\bibinfo {author} {\bibfnamefont {J.}~\bibnamefont
  {Snoek}}, \bibinfo {author} {\bibfnamefont {H.}~\bibnamefont {Larochelle}}, \
  and\ \bibinfo {author} {\bibfnamefont {R.~P.}\ \bibnamefont {Adams}},\ }in\
  \href {https://proceedings.neurips.cc/paper_files/paper/2012/file/
  05311655a15b75fab86956663e1819cd-Paper.pdf} {\emph {\bibinfo {booktitle}
  {Advances in Neural Information Processing Systems}}},\ Vol.~\bibinfo
  {volume} {25},\ \bibinfo {editor} {edited by\ \bibinfo {editor}
  {\bibfnamefont {F.}~\bibnamefont {Pereira}}, \bibinfo {editor} {\bibfnamefont
  {C.}~\bibnamefont {Burges}}, \bibinfo {editor} {\bibfnamefont
  {L.}~\bibnamefont {Bottou}}, \ and\ \bibinfo {editor} {\bibfnamefont
  {K.}~\bibnamefont {Weinberger}}}\ (\bibinfo  {publisher} {Curran Associates,
  Inc.},\ \bibinfo {year} {2012})\BibitemShut {NoStop}%
\bibitem [{\citenamefont {Smith}\ and\ \citenamefont
  {Topin}(2019)}]{superconvergence}%
  \BibitemOpen
  \bibfield  {author} {\bibinfo {author} {\bibfnamefont {L.~N.}\ \bibnamefont
  {Smith}}\ and\ \bibinfo {author} {\bibfnamefont {N.}~\bibnamefont {Topin}},\
  }\bibfield  {booktitle} {\emph {\bibinfo {booktitle} {Artificial Intelligence
  and Machine Learning for Multi-Domain Operations Applications}},\ }\href
  {\doibase 10.1117/12.2520589} {\ \textbf {\bibinfo {volume} {11006}},\
  \bibinfo {pages} {1100612} (\bibinfo {year} {2019})}\BibitemShut {NoStop}%
\bibitem [{\citenamefont {Lasseri}\ \emph {et~al.}(2020)\citenamefont
  {Lasseri}, \citenamefont {Regnier}, \citenamefont {Ebran},\ and\
  \citenamefont {Penon}}]{lasseri2020taming}%
  \BibitemOpen
  \bibfield  {author} {\bibinfo {author} {\bibfnamefont {R.-D.}\ \bibnamefont
  {Lasseri}}, \bibinfo {author} {\bibfnamefont {D.}~\bibnamefont {Regnier}},
  \bibinfo {author} {\bibfnamefont {J.-P.}\ \bibnamefont {Ebran}}, \ and\
  \bibinfo {author} {\bibfnamefont {A.}~\bibnamefont {Penon}},\ }\href
  {\doibase 10.1103/PhysRevLett.124.162502} {\bibfield  {journal} {\bibinfo
  {journal} {Phys. Rev. Lett.}\ }\textbf {\bibinfo {volume} {124}},\ \bibinfo
  {pages} {162502} (\bibinfo {year} {2020})}\BibitemShut {NoStop}%
\bibitem [{\citenamefont {Schran}\ \emph {et~al.}(2020)\citenamefont {Schran},
  \citenamefont {Brezina},\ and\ \citenamefont {Marsalek}}]{committeSchran}%
  \BibitemOpen
  \bibfield  {author} {\bibinfo {author} {\bibfnamefont {C.}~\bibnamefont
  {Schran}}, \bibinfo {author} {\bibfnamefont {K.}~\bibnamefont {Brezina}}, \
  and\ \bibinfo {author} {\bibfnamefont {O.}~\bibnamefont {Marsalek}},\ }\href
  {\doibase 10.1063/5.0016004} {\bibfield  {journal} {\bibinfo  {journal} {The
  Journal of Chemical Physics}\ }\textbf {\bibinfo {volume} {153}} (\bibinfo
  {year} {2020}),\ 10.1063/5.0016004},\ \bibinfo {note} {104105}\BibitemShut
  {NoStop}%
\end{thebibliography}

%

\section{Methods}

\subsection{Dataset Generation and Splitting}
\label{sec:dataset_splitting}

We generated an initial dataset consisting of 200 pairs of Slater determinants, which we randomly partitioned into training, validation, and test sets. Each Slater determinant is represented by its $(330 \times 330)$ one-body density matrix
, generated using a Hartree-Fock code with an \textit{ab-initio} interaction. 
Each data point corresponds to an Oxygen-16 isotope, axially constrained by the $Q_{20}$ collective variable. 
To check the stability of the results to the sampling of the training, validation, and testing sets, we performed 5-fold cross-validation.
Each fold correspond to a different sampling of these data sets while keeping the partition 70\% for the training
set, 20\% for the validation set and 10\% for the test set.
Table \ref{tab:rms_fold} presents the root mean square (RMS) of the Hartree-Fock energy for each fold.

\begin{table}[!h]
\begin{tabular}{l|ccc}
\hline
RMS$(E_{HF})$ (keV) & train (70\%) & valid (20\%) & test (10\%) \\
\hline
Fold 1  & 88 & 291 & 113 \\
Fold 2  & 83 & 287 & 108 \\
Fold 3  & 95 & 298 & 131 \\
Fold 4  & 101 & 299 & 102 \\
Fold 5  & 98 & 308 & 122 \\
\hline
\end{tabular}
\label{tab:rms_fold}
\caption{Root mean square error of the Hartree-Fock energy for 5 different folds of the dataset, for a fixed partition of the train/validation/test sets.
These values are all below 0.3\% of the ground state energy.}
\end{table}

The stability of the method across different dataset splits indicates that our training procedure and evaluation of generalization are suitable for this problem.

\subsection{Variational Autoencoder implementation, architecture and training:}
\label{sec:vae_training}

The non-linear transformation involved in building the new collective variable $\lambda$
relies on a variational autoencoder (VAE)~\cite{kingma2019introduction}.
This type of neural generative model learns a probabilistic mapping between an input data space and a lower-dimensional latent space. 
The VAE is composed of two main components: an encoder, which maps the input data to a latent space, and a decoder, which reconstructs the original data from the latent space representation. 
The primary objective of a VAE is to learn a continuous and smooth latent space, enabling the generation of new data points by sampling the latent space. 
The objective function of the VAE is the sum of a  reconstruction loss and a regularization term, typically the Kullback-Leibler (KL) divergence, which encourages the latent space to have a specific structure, such as following a Gaussian distribution. 
The total loss of the VAE can be written as 
a sum of two terms:
\begin{equation}
\mathcal{L}(\mu, \sigma^2, x, \hat{x}) = \mathcal{L}_{\text{rec.}}(x, \hat{x}) + \mathcal{L}_{\text{KL}}(\mu, \sigma^2).
\end{equation}

The reconstruction loss $\mathcal{L}_{\text{rec.}}(x, \hat{x})$ measures the difference between the original input data $x$ and the reconstructed data $\hat{x}$ generated by the decoder. Assuming Gaussian likelihood with a variance of one in each $x$ dimension the reconstruction loss for a single data point $x$ and its reconstructed output $\hat{x}$ can be expressed as:
\begin{equation}
\mathcal{L}_{\text{rec.}}(x, \hat{x}) = \frac{1}{2}\sum_{i=1}^{D} \left(x_i - \hat{x}_i \right)^2
\end{equation}
where $D$ is the dimension of the data.

The KL divergence term of the loss measures the difference between the approximate posterior distribution $q_{\phi}(z|x)$, represented by the encoder's output, and the true prior distribution $p(z)$, typically assumed to be a standard normal distribution, i.e., $\mathcal{N}(0, I)$. The KL divergence for a single data point can be expressed as:
\begin{equation}
\mathcal{L}_{\text{KL}}(\mu, \sigma^2) = -\frac{1}{2} \sum_{i=1}^{K} \left(1 + \log(\sigma_i^2) - \mu_i^2 - \sigma_i^2 \right),
\end{equation}
where $K$ is the dimension of the latent space, $\mu_i$ is the mean of the approximate posterior distribution for the $i$-th latent variable, and $\sigma_i^2$ is its variance.

Our VAE was implemented using PyTorch and the training process was completed in around 10k epochs, taking approximately 20 minutes on a single RTX 3090 GPU. Early stopping was implemented based on the validation set.

The VAE architecture was optimized through Bayesian optimization of its hyperparameters~\cite{NIPS2012_05311655}, including the number of hidden layers, density of each layer, and the optimizer. 
To assess the quality of the reconstruction, the mean squared error of the energy on the validation set was used. To accelerate the training procedure and consequently reduce the time footprint of the Bayesian optimization algorithm, we employed an implementation of the superconvergence method \cite{superconvergence} to accelerate the training. Superconvergence is a phenomenon observed in the training of deep learning models, where the convergence rate during optimization is significantly accelerated compared to traditional training methods. This rapid convergence is achieved by tuning hyperparameters such as the learning rate, batch size, and weight decay, along with employing techniques like cyclical learning rate schedules, learning rate warm-up, and regularization methods.

In our case, we used a cyclical learning rate, which adjusts the learning rate periodically throughout training, alternating between higher and lower values. This approach allows the model to escape local minima and explore the loss landscape more effectively, leading to faster convergence and improved generalization. We also used the learning rate warm-up technique, where the learning rate is gradually increased during the initial phase of training, allowing the model to adapt more smoothly to the optimization process. No additional regularization was employed.

The final architecture consists of symmetric Encoder and Decoder, with their weights trained independently from each other, and four hidden layers each.

\begin{itemize}
\item \textbf{Encoder:}
\begin{enumerate}
\item Input layer: 20 neurons
\item Hidden layer 1: 100 neurons
\item Hidden layer 2: 80 neurons
\item Hidden layer 3: 60 neurons
\item Hidden layer 4: 10 neurons
\item Latent space layer ($\lambda$): 1 neuron
\end{enumerate}
\item \textbf{Decoder:}
\begin{enumerate}
\item Hidden layer 1: 10 neurons
\item Hidden layer 2: 60 neurons
\item Hidden layer 3: 80 neurons
\item Hidden layer 4: 100 neurons
\item Output layer: 20 neurons
\end{enumerate}
\end{itemize}


\subsection{Uncertainty of the potential barrier height}
\label{sec:heights_sensitivity}

One way to check the robustness of our approach is to use the concept of committees of neural networks. 
The idea described for instance in \cite{lasseri2020taming} and \cite{committeSchran} is to clone several time the neural model. 
The clones are identical in terms of architecture and the only difference in between the members of this neural committee consists on the initial weights initialization. 
Theses differences in the initial weights allow the probing of the variational space explored during the training phase for a fixed dataset. 
Each model of the committee will then have a prediction on its own. 
If the modelisation is robust enough the predictions will be relatively close to each other, if it is not the case one can see this result as an hint that the specific datapoint predicted is an outlayer. 
Moreover in the first case the spread of the predictions can be directly linked to the uncertainty of the prediction and as such is a good tool to estimate the reliability of the model.
Our specific case is quite interesting: from a physical point of view the topology of the latent space does not matter at all since it describe a non-observable representation. 
The only characteristic that can be used to quantify the performance of our approach is the one which is directly linked to an experimental observable e.g. the height of the energy barrier predicted. 
As such by using a committee of VAE, one can probe the sensitivity and robustness of our model. 
In our case for a committee of 10 VAEs we can estimate the height of the barrier by computing the mean and the variance over the different VAE, obtaining $E_{barrier} = 13.21 \pm 2.10$ MeV as such the results are coherent with each other and reinforce the robustness claim of our approach.
\begin{figure}[!ht]
\includegraphics[width=0.5\textwidth]{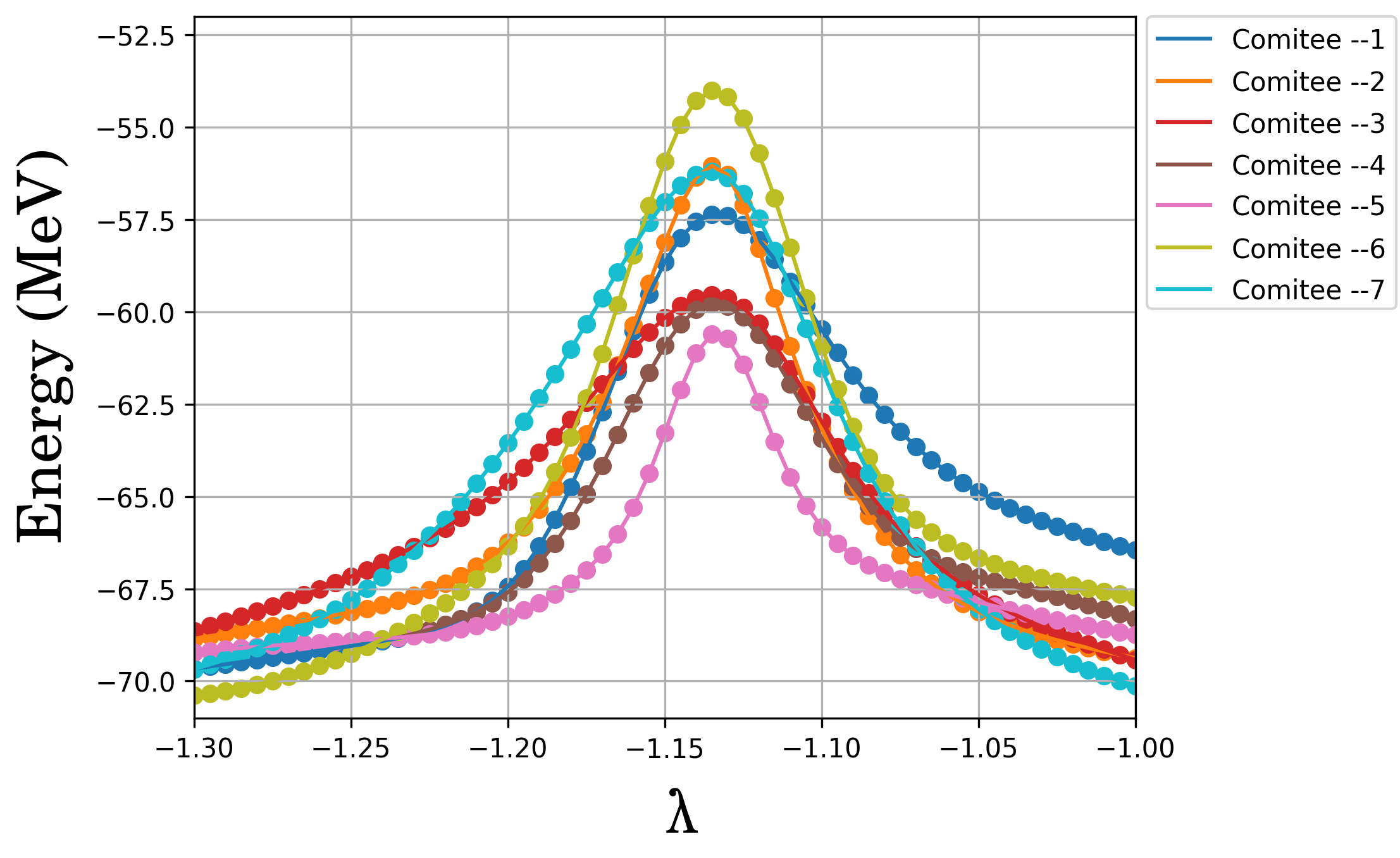}
\caption{For a committee of 10 VAEs, (7 displayed here) the sampling of the latent space at the level of the discontinuity. The presence of a new potential barrier
is a robust feature while the height this barrier varies from a member of the committee to another. The standard deviation of this variation is 2.10 MeV which
represents 16\% of its mean value.}
\label{fig:barrier_comittee}
\end{figure}

\section{Acknowledgements}

The authors would like to thank S. Hilaire for his contribution in testing this approach with a relativistic energy density functional structure code, 
J.-P. Ebran for his insight on relativist energy density functional formalism, N.~Dubray for his support with the \texttt{HFB3} solver and 
A. Penon for his advices on training the VAE.
This work was partly performed under the auspices of the U.S. Department of Energy by Lawrence Livermore National Laboratory under Contract No. DE-AC52-07NA27344 and by the Office of LDRD.

\section{Author contributions}

R.-D.~Lasseri and D. Regnier developed and implemented the core ideas to build new collective variables from a set of existing Slater determinants.
M. Frosini contributed to the production of the \textit{ab-inito} Hartree-Fock series of $^{16}$O states.
M.~Verriere and N.~Schunck provided technical consulting 
on the parameteization of Slater determinant states for use with neural networks.
D.~Regnier and R.~Lasseri wrote the first version of this article that was further read, amended, corrected and approved by all authors.

\end{document}